\documentclass[twocolumn]{aastex631}
\usepackage{amsmath}
\usepackage{amssymb}
\usepackage{array}
\usepackage{float}
\usepackage{tabularx}
\usepackage{xcolor}
\usepackage{ulem}
\usepackage{hyperref}
\usepackage{multirow}
\usepackage{booktabs}

\begin{document}

\title{A Rare Population of Intermediate-mass Helium Stars Between Hot Subdwarfs and Wolf-Rayet Stars}
\author[0009-0002-3654-8775]{Gui-Yu Wang}
\affiliation{Department of Astronomy, Nanjing University, Nanjing 210023, People's Republic of China}
\affiliation{Key Laboratory of Modern Astronomy and Astrophysics, Nanjing University, Ministry of Education, Nanjing 210023, People's Republic of China}

\author[0000-0003-2506-6906]{Yong Shao}
\email{shaoyong@nju.edu.cn}
\affiliation{Department of Astronomy, Nanjing University, Nanjing 210023, People's Republic of China}
\affiliation{Key Laboratory of Modern Astronomy and Astrophysics, Nanjing University, Ministry of Education, Nanjing 210023, People's Republic of China}

\author[0000-0003-3862-0726]{Jian-Guo He}
\affiliation{Department of Astronomy, Nanjing University, Nanjing 210023, People's Republic of China}
\affiliation{Key Laboratory of Modern Astronomy and Astrophysics, Nanjing University, Ministry of Education, Nanjing 210023, People's Republic of China}

\author[0009-0009-4482-6350]{Yu-Dong Nie}
\affiliation{Department of Astronomy, Nanjing University, Nanjing 210023, People's Republic of China}
\affiliation{Key Laboratory of Modern Astronomy and Astrophysics, Nanjing University, Ministry of Education, Nanjing 210023, People's Republic of China}

\author[0000-0002-3614-1070]{Xiao-Jie Xu}
\affiliation{Department of Astronomy, Nanjing University, Nanjing 210023, People's Republic of China}
\affiliation{Key Laboratory of Modern Astronomy and Astrophysics, Nanjing University, Ministry of Education, Nanjing 210023, People's Republic of China}

\author[0000-0002-0584-8145]{Xiang-Dong Li}
\affiliation{Department of Astronomy, Nanjing University, Nanjing 210023, People's Republic of China}
\affiliation{Key Laboratory of Modern Astronomy and Astrophysics, Nanjing University, Ministry of Education, Nanjing 210023, People's Republic of China}

\begin{abstract}
Helium stars stripped of their hydrogen envelopes represent pivotal phases in binary evolution, yet their origins, particularly within the intermediate-mass range of $2-8\, M_{\odot}$, still remain poorly understood. This population bridges the gap between low-mass hot subdwarfs and massive Wolf-Rayet stars, but has remained largely unobserved. In this study, we employ binary population synthesis to systematically investigate the formation and properties of intermediate-mass helium stars (IMHeS) across various galactic metallicities. Our results indicate that metallicity and common-envelope ejection efficiency are the dominant factors shaping the IMHeS population. We estimate that several thousand IMHeS exist in the Milky Way, with several hundred more in the Magellanic Clouds. The vast majority of IMHeS reside in binaries, with fewer than $10\%$ appearing as single stars. Among IMHeS binaries, $\gtrsim 50\%$ are expected to have main-sequence companions, and the remainder host compact companions (including helium stars, white dwarfs, neutron stars, or black holes). The former systems form mainly through stable mass transfer, whereas the latter arise predominantly from common envelope evolution. 
Our work provides quantitative predictions for the populations of these elusive stars formed through binary interactions and offers guidance for future observational searches.
\end{abstract}


\section{Introduction} \label{sec: introduction}

Helium (He) stars are the evolved remnants of post main-sequence (MS) stars that have lost their hydrogen-rich envelopes through intense stellar winds or binary interactions \citep{Han2002,shenar2024wolfrayetstars}. 
Historically, He stars have been classified into two distinct groups: hot subdwarf (sdOB) stars with masses below $\sim 2 \,M_{\odot}$ \citep{Heber2009}, and Wolf-Rayet (WR) stars with masses exceeding $\sim 8 \,M_{\odot}$ \citep{Crowther2007}. 
The latter classification, however, is observationally motivated, and the minimum mass at which a stripped star exhibits a WR-type spectrum is known to depend on metallicity \citep{Shenar2020A&A...639L...6S}.
Current estimates suggest the presence of approximately 2000 WR stars in the Milky Way \citep{shenar2024wolfrayetstars}, while more than 6000 hot subdwarf stars have been detected within 5 kpc of the Sun \citep{Geier2020A&A...635A.193G}, with over 60000 additional candidates identified by \cite{Culpan2022A&A...662A..40C} from Gaia DR3 \citep{GaiaDataRelease3_2023}. Between these two well-studied populations, there is an obvious gap, i.e., a scarcity of He stars in the intermediate-mass range of $2-8\, M_{\odot}$. 
 
In recent years, however, detections of such IMHeS have been on the rise. For instance, \citet{Irrgang2022} identified the stripped pulsating core star $\gamma$ Columbae, with a mass of $4 - 5\, M_{\odot}$. \cite{Villase2023MNRAS.525.5121V} proposed that VFTS 291 contains a recently stripped star of $\sim 1.5-2.5\,M_{\odot}$ formed via mass transfer (MT). 
By combining ultraviolet photometry with optical spectroscopy, \citet{Drout2023Sci...382.1287D} discovered 25 sources in the Large and Small Magellanic Clouds (LMC/SMC) consistent with theoretical predictions for binaries hosting IMHeS, with 10 systems spectroscopically analyzed by \cite{Gotberg2023ApJ...959..125G}. In the binary system HD 45166, \cite{Shenar2023Sci...381..761S} studied a magnetic quasi-WR companion and constrained its mass to $2.03 \pm 0.44\,M_{\odot}$, suggesting a possible origin from the merger of two lower-mass He stars. Similarly, \cite{Ramachandran2023A&A...674L..12R} identified a $\sim 3\, M_{\odot}$ He star in the binary SMCSGS-FS 69 with a Be-star companion.

IMHeS are recognized as the progenitors of various types of stripped-envelope supernovae (SNe), including core collapse SNe \citep{Zapartas2017ApJ...842..125Z,Zapartas2026,Ercolino2026} and electron capture SNe \citep{Podsiadlowski2004ApJ...612.1044P,Poelarends2017ApJ...850..197P,Guo2024MNRAS.530.4461G}, that arise from binary interactions. 
In compact binaries, they may also give rise to ultra-stripped SNe \citep{Tauris2015MNRAS.451.2123T}. If the binary survives the SN explosion, these systems can evolve into double neutron stars (NSs) and become sources of gravitational waves \citep[e.g.,][]{Tauris2017ApJ...846..170T, K.Dedoi:10.1126/science.aas8693}. 
Moreover, IMHeS have been proposed as significant sources of ionizing photons in stellar systems, playing a crucial role in the ionization of interstellar media and the photometric evolution of galaxies \citep{Dionne2006ApJ...641..252D, Gotberg2018A&A...615A..78G, Doughty2021MNRAS.505.2207D}.
Overall, IMHeS serve as critical evolutionary links connecting massive binary systems with initial component masses of $\sim 8-25\,M_\odot$ to their final fates \citep{Drout2023Sci...382.1287D}, including compact object (CO) binaries and various SN types.

Recent studies have begun to explore the formation and evolution of IMHeS. Binary interactions are critical in determining whether and when Roche lobe overflow (RLOF) or common-envelope (CE) evolution occurs, thereby shaping the final fate of IMHeS. By using MESA, \cite{Yungelson2024A&A...683A..37Y} traced the evolution of Galactic He stars formed via stable RLOF and estimated $\sim 20000$ He+MS systems with $M_{\mathrm{He}}> 1\,M_{\odot}$ and $\sim 3000$ with $M_{\mathrm{He}} > 2\,M_{\odot}$. In addition, \cite{Hovis-Afflerbach2025A&A...697A.239H} highlighted the metallicity dependence of the mass distribution of hot stripped stars and predicted $\sim 30000$ systems with He stars above $1\,M_{\odot}$ and $\sim 4000$ above $2.6\,M_{\odot}$, for a constant star formation rate (SFR) of 1 $M_{\odot}\,\mathrm{yr^{-1}}$ and regardless of metallicity. 
Observationally, \citet{Blomberg2026} predicted populations of stripped stars in the $1-10\, M_{\odot}$ range to be around 1000 in the SMC and 2500 in the LMC, though these estimates are subject to uncertainties due to dust extinction, contamination from bright companions, and incomplete survey coverage. Besides MS stars \citep[see also][]{Shao2021a,Wang2024,Dutta2024,Xu2025,Schurmann2025}, the companion of IMHeS could also be a black hole \citep[BH,][]{Shao2020,Sen2025}, an NS \citep{shao2019,Li2024}, or a white dwarf \citep[WD,][]{Wang2009,Cui2022RAA....22b5001C}.

Despite these advances, a systematic investigation into the formation channels and population statistics of IMHeS, particularly under varying environmental conditions and binary evolution models, remains lacking. 
In this study, we employ binary population synthesis to comprehensively explore the properties of IMHeS, incorporating a range of physical models for metallicity, CE evolution, MT efficiency, and stellar winds. 

The structure of this paper is organized as follows. Section \ref{sec: method} outlines the physical models employed in our study. Section \ref{sec: result} presents the binary population synthesis results, detailing the evolutionary channels and properties of IMHeS populations. Finally, we discuss our findings and present conclusions in Section \ref{sec: conclusion}. 

In some figures of this paper, we utilized the \texttt{kdeplot} method from the \texttt{seaborn} library \citep{Waskom2021zndo....592845W} to visualize the contour distributions of physical parameters. This approach generates smooth contours that effectively represent the underlying probability density of the data in a continuous, two-dimensional form. A number threshold of $10^{-5}$ was applied to exclude low-probability system, ensuring that only statistically significant sources were included in the analysis and subsequent visualization.

\section{Method} \label{sec: method}
\subsection{Stellar evolution code} \label{sec: BPS}
We employ the binary star evolution code \texttt{BSE} to simulate the formation and evolutionary pathways of IMHeS. The original \texttt{BSE} framework was developed by \citet{Hurley2002} and later refined by \citet{Kiel2006MNRAS.369.1152K}. Our study utilizes a version significantly modified by \citet{Shao2014}, which includes improved treatments for MT stability and stellar wind processes. Additionally, \citet{Shao2021b} integrated multiple SN explosion mechanisms involving the rapid \citep{Fryer2012}, delayed \citep{Fryer2012}, and stochastic \citep{Mandel2020MNRAS.499.3214M} prescriptions to account for the mass distribution of the resulting compact remnants. In the present work, we adopt the delayed SN mechanism, motivated by recent detections of COs in the mass gap between NSs and BHs \citep{Shao2022,Wang2024NatAs...8.1583W,Abac2024}. 

\subsection{Physical Assumptions} \label{sec: PA}

Our models are built upon several key input parameters that govern the formation and evolution of IMHeS. These include metallicity, CE ejection efficiency, MT efficiency, and stellar wind prescriptions for He stars. The specific implementations are as follows.

\textit{Metallicity.} We consider four representative metallicity ($Z$) values. The default model adopts $Z = Z_{\odot}$ for Milky Way binaries, where $Z_{\odot}=0.02$. Additional models include $1/2\, Z_{\odot}$ for the LMC and $ 1/4 \,Z_{\odot}$ for the SMC, and $Z = 1/20\, Z_{\odot}$ for an extremely low-metallicity environment.

\textit{MT efficiency.} We implement three MT modes from \cite{Shao2014} to deal with the evolution of primordial binaries consisting of a massive primary and a less-massive secondary. The rotation-dependent mode (MT1) links the accretion rate to the secondary's rotational velocity, ceasing accretion once the secondary reaches its critical rotational velocity. The half-mass accretion mode (MT2) assumes half of the transferred mass is accreted by the secondary, with the remainder lost from the system. The thermal-equilibrium-limited mode (MT3) restricts the accretion rate based on the secondary's thermal timescale, approximating conservative MT. In all modes, material lost from the binary system is assumed to carry specific angular momentum of the secondary.

\textit{CE evolution.} Dynamically unstable MT leads to CE evolution. We adopt the standard energy formalism  \citep{Webbink1984} and use the binding energy parameter from \citet{Xu2010} and \citet{Wangchen2016RAA} to deal with the orbital evolution of CE binaries. We test three values for the CE ejection efficiency with $\alpha_{\mathrm{CE}} = 1$, 3, and 5. A value of $\alpha_{\mathrm{CE}} = 1$ implies that all orbital energy is used to eject the envelope \citep[e.g.,][]{Zuo2014MNRAS.442.1980Z}, while higher values allow for the formation of wider post-CE binaries and are necessary to explain observed systems such as IK Peg \citep{Davis2010MNRAS.403..179D} and IC 10 X-1 \citep{Wang2024ApJ...974..184W}. In some cases, values as high as $\alpha_{\mathrm{CE}} \gtrsim 5$ have been suggested \citep{Fragos2019,Deng2024}.

\textit{Stellar wind.} Wind mass loss is implemented following \cite{Belczynski2010ApJ...714.1217B}. For stripped He stars,  \cite{Vink2017A&A...607L...8V} suggested mass-loss rates approximately an order of magnitude lower than those extrapolated from Wolf-Rayet star calibrations \citep[see also][]{Gotberg2023ApJ...959..125G}. Accordingly, we explore two scaling factors for He star winds, i.e., $\eta_{\mathrm{He}}=1$ (standard) or $\eta_{\mathrm{He}}=0.1$ (reduced). 

A total of nine distinct models are constructed to explore the effects of different physical assumptions, as summarized in Table 1. Our fiducial model (Model A) adopts solar metallicity ($Z=Z_\odot$), a CE ejection efficiency of $\alpha_{\mathrm{CE}} = 1$, the rotation-dependent MT scheme, and a He wind scaling factor of $\eta_{\mathrm{He}}=1$. To compare with the fiducial model, variations include different metallicities (Models B/C/D), alternative MT schemes (Models E/F), different $\alpha_{\mathrm{CE}}$ values (Models G/H), and a reduced wind scaling factor with $\eta_{\mathrm{He}}=0.1$ (Model I).

\subsection{Population Synthesis Method} \label{sec: PB}

We simulate the evolution of $10^6$ primordial binaries for each model, each consisting of two zero-age main-sequence (ZAMS) stars formed simultaneously on circular orbits. The initial primary mass $M_{\mathrm{1,i}}$ ranges from 5 $M_{\odot}$ to 100 $M_{\odot}$, the initial secondary mass $M_{\mathrm{2,i}}$ from 0.1 $M_{\odot}$ to 100 $M_{\odot}$, and the initial orbital separation from 3 $R_{\odot}$ to 10000 $R_{\odot}$. This large sample serves as a representative distribution of all binary systems in a galaxy. Each simulated binary is evolved from ZAMS to its final fate, and its contribution to the IMHeS population is recorded over time. 

We assume a binary fraction of unity, meaning that all stars are initially born in binary systems. Observations indicate that the majority of intermediate- and high-mass stars are found in binaries \citep[e.g.,][]{Offner2023}, although the binary fraction is mass-dependent and below unity. This assumption may lead to an overestimate of the predicted IMHeS population by a factor of less than 2.

We adopt a constant star formation rate (SFR) of 1 $M_{\odot}\,\mathrm{yr^{-1}}$ over a period of 10 Gyr for all models. This timescale ensures that the simulated stellar population reaches a steady state, in which the birth and death rates of IMHeS are balanced. Specifically, following the population synthesis methodology of \citet{Shao2021b}, we sum over all timesteps in which a system contains an IMHeS, and multiply by the SFR to obtain the expected number in a galactic environment. The impact of galaxy-specific SFRs is discussed in Section \ref{sec: conclusion}.

\section{Result} \label{sec: result} 

\begin{table*}[htbp]
    \centering
    \caption{Predicted numbers of IMHeS under different physical models, by assuming a constant SFR of $1\,M_{\odot}\,\mathrm{yr^{-1}}$. The populations are categorized by both their formation channel (Case A MT, Case B MT, or CE evolution) and the type of companion star. The total number for each model is given in the rightmost column.}
    \begin{tabular}{l|l|l|ccccccc|c}
    \toprule 
     & Models & Channels & $N_{\mathrm{He+MS}}$ & $N_{\mathrm{He+G}}$ & $N_{\mathrm{He+He}}$ & $N_{\mathrm{He+WD}}$ & $N_{\mathrm{He+NS}}$ & $N_{\mathrm{He+BH}}$ & $N_{\mathrm{Single\,IMHeS}}$ & $N_{\mathrm{total}}$\\ 
    \midrule 
    \cline{1-11}
    \multirow{4}{*}{$\textbf{Fiducial}$} 
    & $\textbf{Model A}$    & Case A & 100   & $<1$    & 2.1     & 0    & 0.8  & 12.6  & $<1$  & 115.5 \\
    & Z=0.02 MT1            & Case B & 504   & 7.3     & 1.9     & 0    & 2    & 28.4  & $<1$  & 543.6 \\
    & $\alpha_{\mathrm{CE}}=1$   & CE     & 350   & 2.0     & 49.7    & 30.6 & 21.4 & 22.1  & 42.8  & 518.6 \\
    & $\eta_{\mathrm{He}}=1$     & All    & 954   & 10      & 54      & 30.6 & 24.2 & 63    & 44.3  & 1178 \\
    \cline{1-11}
    \multirow{12}{*}{Metallicity}
    & $\textbf{Model B}$    & Case A & 87   & $<1$    & 1.5     & 0    & 0.3  & 7.6   & 1.1   & 97.5   \\ 
    & Z=0.01                & Case B & 482   & 7.7     & $<1$    & 0    & 1.3  & 28.2  & $<1$  & 519.2 \\
    &                       & CE     & 335   & 1.1     & 49      & 40.7 & 20.7 & 18.3  & 62    & 526.8 \\
    &                       & All    & 904   & 9       & 51      & 40.7 & 22.3 & 54    & 64    & 1145 \\
    \cline{2-11}
    & $\textbf{Model C}$    & Case A & 81    & $<1$    & 1       & 0    & 0.08 & 3.7   & $<1$  & 85.8 \\
    & Z=0.005               & Case B & 359   & 5.5     & $\sim0$ & 0.01 & 0.24 & 20.2  & 2.6   & 387.5 \\
    &                       & CE     & 78    & $<1$    & 13      & 15.9 & 8.7  & 5.7   & 22    & 143.3 \\
    &                       & All    & 518   & 6       & 14      & 16   & 9    & 30    & 25    & 618 \\
    \cline{2-11}
    & $\textbf{Model D}$    & Case A & 74    & $<1$    & $<1$    & 0    & 0    & 1.5   & $<1$  & 75.5 \\
    & Z=0.001               & Case B & 351   & 4.8     & $\sim0$ & 0    & 0.1  & 19.1  & 2.9   & 377.9 \\
    &                       & CE     & 259   & $<1$    & 44      & 55.2 & 22.2 & 28.2  & 71    & 479.6 \\
    &                       & All    & 685   & 5.5     & 44      & 55.2 & 22.3 & 49    & 75    & 935 \\
    \cline{1-11}
    \multirow{8}{*}{MT efficiency}
    & $\textbf{Model E}$    & Case A & 91   & $\sim0$ & 2.5     & 0    & 0.3  & 7.5   & $<1$  & 101.3   \\
    & MT2                   & Case B & 507   & $\sim0$ & $\sim0$ & 0    & 0    & 0.3   & 5.4   & 512.7 \\
    &                       & CE     & 353   & 2.2     & 163.6   & 252  & 34.5 & 14.9  & 144.7 & 965.9 \\
    &                       & All    & 951   & 2.3     & 166.1   & 252  & 34.8 & 22.7  & 150.4 & 1579 \\
    \cline{2-11}
    & $\textbf{Model F}$    & Case A & 95    & $\sim0$ & 2.5     & 0    & 0.2  & 7.1   & $<1$  & 104.8 \\
    & MT3                   & Case B & 514   & $\sim0$ & $\sim0$ & 0    & 0    & 0.4   & 4.3   & 518.7 \\
    &                       & CE     & 351   & 2.1     & 160.3   & 255  & 31.3 & 15.4  & 146.2 & 961.3 \\
    &                       & All    & 960   & 2.2     & 162.8   & 255  & 31.5 & 23    & 151.3 & 1586 \\
    \cline{1-11}
    \multirow{8}{*}{CE evolution}     
    & $\textbf{Model G}$    & Case A & 95    & $\sim0$ & 2.3     & 0    & 0.2  & 7.3   & $<1$  & 104.8 \\
    & $\alpha_{\mathrm{CE}}=3$   & Case B & 499   & $\sim0$ & $\sim0$ & 0    & 0    & 0.3   & $<1$  & 499.3 \\
    &                       & CE     & 566   & 7.5     & 370.8   & 554  & 52.1 & 32.3  & 78.8  & 1661.5 \\
    &                       & All    & 1160  & 7.6     & 373.1   & 554  & 52.3 & 40  & 80    & 2267 \\
    \cline{2-11}
    & $\textbf{Model H}$    & Case A & 96    & $\sim0$ & 2.2     & 0    & 0.3  & 7.1   & $<1$  & 105.6 \\
    & $\alpha_{\mathrm{CE}}=5$   & Case B & 456   & $\sim0$ & $\sim0$ & 0    & 0    & 0.3   & $<1$  & 456.3 \\
    &                       & CE     & 560   & 8.8     & 408.4   & 624  & 80.7 & 38.9  & 35.7  & 1756.5 \\
    &                       & All    & 1112  & 8.9     & 410.6   & 624  & 81   & 46.3  & 37    & 2319 \\
    \cline{1-11}
    \multirow{4}{*}{Stellar wind}
    & $\textbf{Model I}$    & Case A & 90    & $\sim0$ & 2.2  & $\sim0$    & 0.2  & 11  & 0.3   & 103.7 \\
    & $\eta_{\mathrm{He}}=0.1$   & Case B & 498   & $\sim0$ & 1.7 & 0    & $\sim0$    & 19.7  & 1.1   & 521.5 \\
    &                       & CE     & 344   & 2.2     & 42.1   & 29  & 24   & 24.6 & 42   & 507.8 \\
    &                       & All    & 932   & 2.3     & 46   & 29  & 24.2 & 55.3 & 43.5 & 1133 \\
    \cline{1-11}
    \end{tabular}
    \label{tab: Num_He}
\end{table*}

\subsection{Impact of Physical Parameters on IMHeS Populations} \label{sec: IPPIP}

\begin{figure*}[htp]
    \centering
    \includegraphics[width=\textwidth]{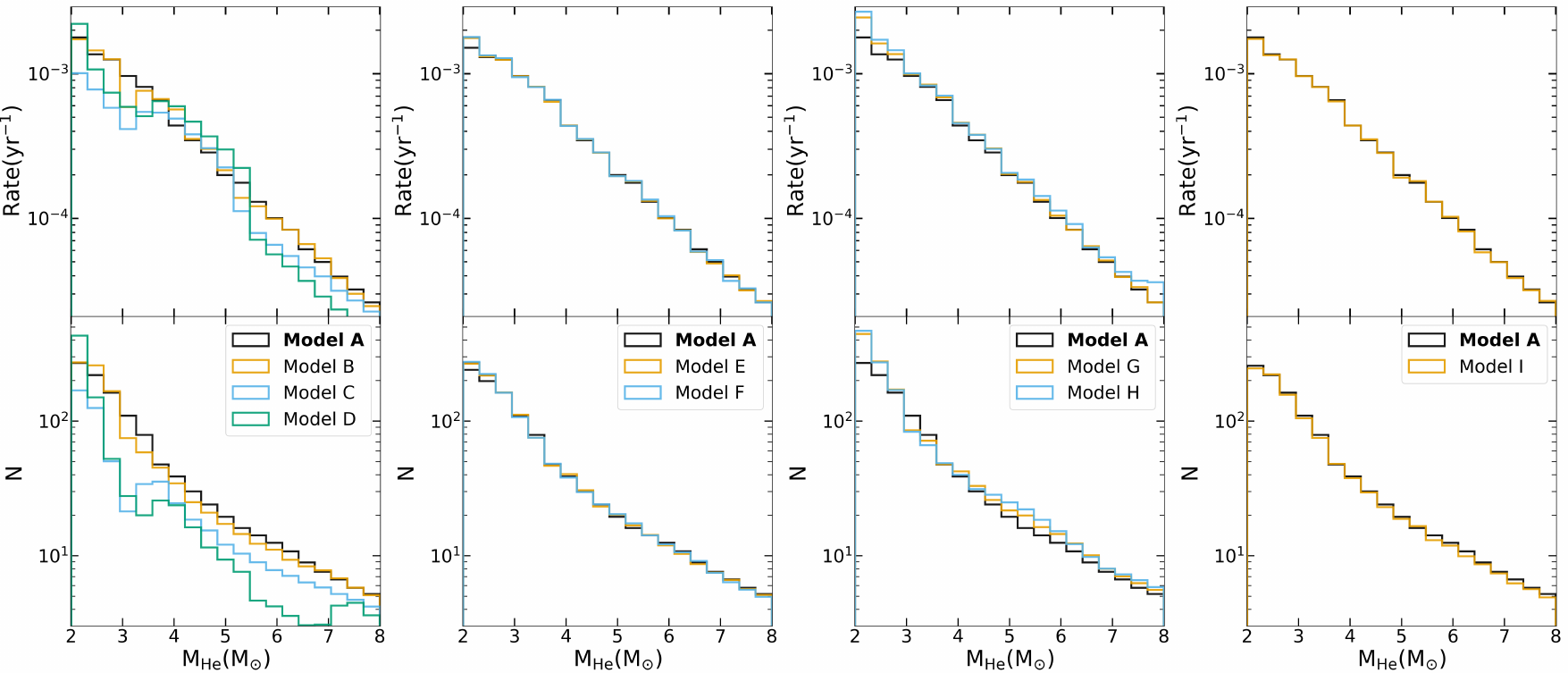}
    \caption{Predicted birthrates (top panel) and total numbers (bottom panel) of IMHeS as a function of $M_{\mathrm{He}}$ for different physical assumptions (assuming a constant SFR of $1\,M_{\odot}\,\mathrm{yr^{-1}}$). From left to right, the panels show results for variations in metallicity, MT efficiency, CE ejection efficiency $\alpha_{\mathrm{CE}}$, and He-star wind scaling factor $\eta_{\mathrm{He}}$. See Table \ref{tab: Num_He} for model configurations.}
    \label{fig: merge_massdistribution_N_rate}
\end{figure*}

We investigate how variations in the four physical parameters (metallicity, CE ejection efficiency, MT efficiency, and He star wind scaling factor) affect the population statistics of IMHeS. Figure \ref{fig: merge_massdistribution_N_rate} illustrates the effects of these parameters on the numbers and birthrates of the IMHeS population as a function of their masses, while Table \ref{tab: Num_He} summarizes the predicted numbers for each model configuration.

A common trend across all models is that both the birthrate and total number of IMHeS decline rapidly with increasing He star mass. Relative to the fiducial model (Model A, $Z=0.02$), the IMHeS population decreases by approximately 3\% (Model B, $Z=0.01$), 48\% (Model C, $Z=0.005$), and 21\% (Model D, $Z=0.001$). 
This suppression arises from two related mechanisms. Lower metallicity reduces stellar wind mass loss, leading to more compact stellar structures that delay the onset of RLOF. Consequently, when RLOF occurs, the donor star has evolved to a later stage, favoring CE evolution over stable mass transfer (SMT). Such evolution often leads to mergers, thereby reducing the number of IMHeS that ultimately form.

Comparing the three MT schemes (Models A, E, and F, corresponding to MT1, MT2, and MT3, respectively), the total IMHeS number increases from $\sim1180$ (Model A) to $\sim1580$ (Model E) and $\sim 1590$ (Model F) $-$ enhancements of approximately 34\% and 35\%, respectively.  
This increase is primarily due to higher MT efficiency, which allows the secondary star to accrete more mass. The resulting growth of the He-core mass enables some systems that would otherwise host lower-mass He stars to instead evolve into IMHeS binaries.

For CE ejection efficiency, increasing $\alpha_{\mathrm{CE}}$ from 1 to 3 (Model G) raises the total IMHeS population to $\sim2270$, nearly doubling that of the fiducial model. A further increase to $\alpha_{\mathrm{CE}}=5$ (Model H) yields a modest additional increase to $\sim2320$. This trend reflects the fact that higher $\alpha_{\mathrm{CE}}$ values make it easier to eject the CE, thereby preventing mergers and allowing more binaries to survive as IMHeS systems.

In contrast, reducing the He wind scaling factor from $\eta_{\mathrm{He}}=1$ to $\eta_{\mathrm{He}}=0.1$ (Model I) results in a total IMHeS population of $\sim 1130$, which is nearly identical to that of the fiducial model. This is because the majority of IMHeS already experience sufficiently weak winds at $\eta_{\mathrm{He}}=1$; therefore, a further reduction by an order of magnitude has only a negligible effect on their evolution.

\subsection{Formation Channels and Properties of IMHeS} \label{sec: PBI}

We classify the formation channels of IMHeS according to the evolutionary states of their direct progenitors into three primary categories: Case A MT, Case B MT, and CE evolution. 

The Case A MT channel involves SMT occurring while the progenitor of IMHeS is still on the MS. The Case B MT channel denotes SMT that begins after the progenitor has ended its MS evolution but before the onset of core He burning. The CE evolution channel involves systems where the progenitor undergoes dynamically unstable MT, leading to a CE phase. 
Systems undergoing multiple MT episodes are labeled sequentially (e.g., $\mathrm{SMT+SMT, SMT+CE, CE+CE}$). Other formation channels are omitted due to their negligible contributions. 

To construct our IMHeS  sample, we select systems that have undergone binary interactions. IMHeS formed via SMT include He stars in the contraction phase with a thin hydrogen envelope (objects flagged as giant stars in the \texttt{BSE} code and selected when they have been sufficiently stripped such that the binary becomes detached) and subsequent evolutionary phases, while those from CE evolution are treated as naked He stars, consistent with the default implementation in the \texttt{BSE} code \citep{Hurley2002}.

As summarized in Table \ref{tab: Num_He}, under the fiducial model (Model A), IMHeS form predominantly via the Case B MT and CE evolution channels, contributing $\sim 1060$ systems. An additional population of $\sim 120$ IMHeS originates from the Case A MT channel. 

We further analyze the primary formation channels of IMHeS and explore the distribution of relevant parameters within the fiducial model. The IMHeS populations are classified into five categories based on companion type:

(1) He+MS binaries. IMHeS with an MS companion.

(2) He+G binaries. IMHeS with a giant companion.

(3) He+He binaries. Systems consisting of two He stars, at least one of which is in the intermediate-mass range.

(4) He+CO binaries. IMHeS with a CO companion, which can be a WD, an NS, or a BH.

(5) Single IMHeS. Isolated IMHeS that have lost their hydrogen envelopes via binary interactions.

\subsubsection{He+MS Binaries} \label{sec: res_HeMS}
He+MS binaries represent the dominant population of IMHeS and may serve as progenitors for other types of IMHeS binaries, such as He+He systems. 

\begin{figure}[htbp]
\centering
\includegraphics[width=0.4\textwidth]{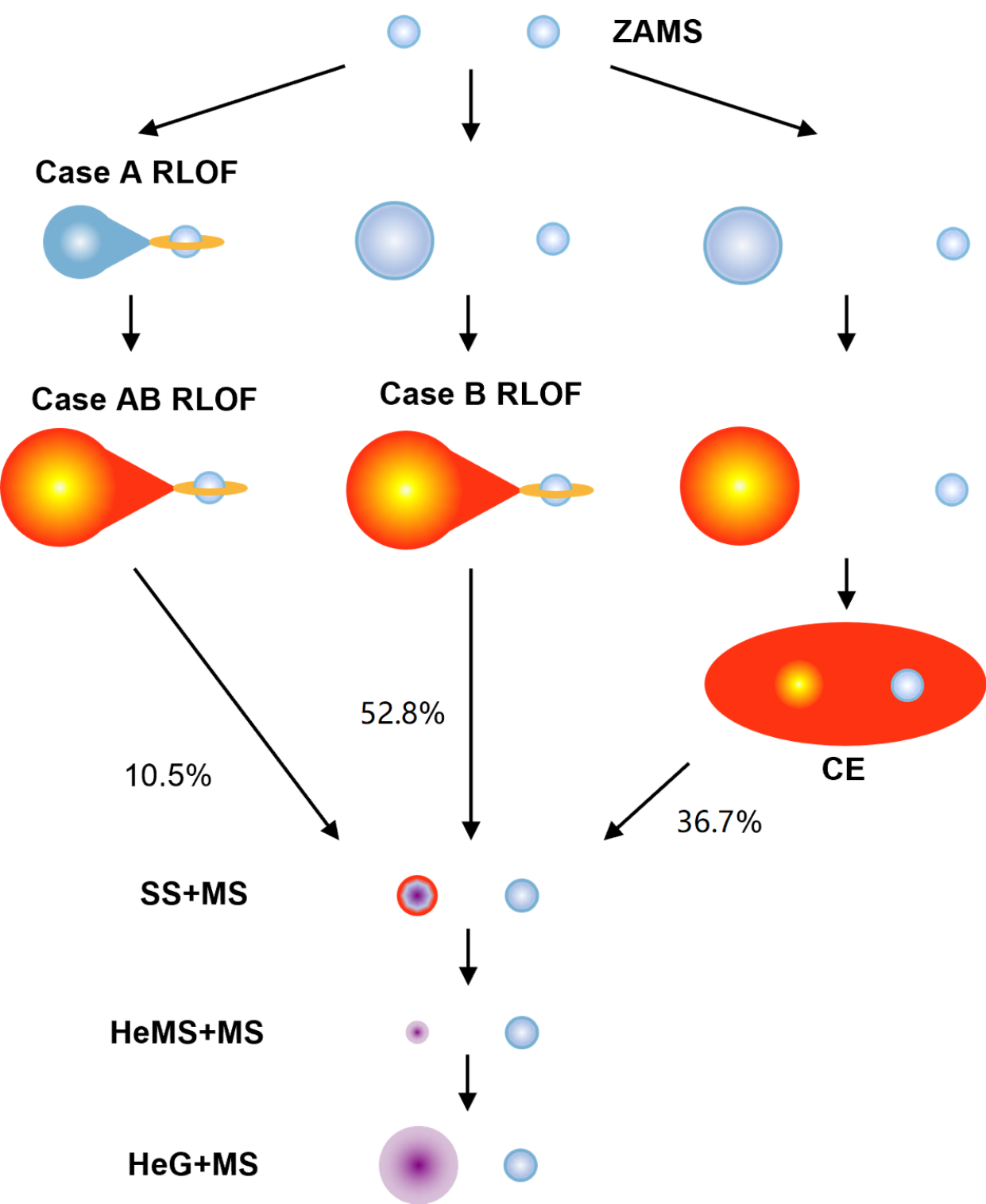}

\caption{Schematic illustration of the three primary evolutionary channels leading to the formation of He+MS binaries. The normalized number fractions of IMHeS formed via each channel in our fiducial model are marked, with Case B MT and CE evolution being the dominant pathways. In each channel, the primordial binary undergoes RLOF, resulting in the stripping of the primary's hydrogen envelope and the formation of a stripped star (SS). The SS subsequently evolves into a He main-sequence (HeMS) star and later into a He giant (HeG), while the secondary remains on the MS stage, forming the He+MS binary system.}
\label{fig: HeMS_evol}
\end{figure}

Figure \ref{fig: HeMS_evol} illustrates the three primary formation channels of He+MS binaries. In the Case A MT channel, primordial binaries with short orbital periods undergo long-timescale RLOF until the primary finishes its MS evolution.
Once the primary's hydrogen envelope is stripped, it becomes a stripped star (SS), forming an SS+MS system. After the SS loses its remaining hydrogen envelope via stellar winds, a HeMS or HeG star is formed alongside an MS companion. In the Case B MT channel, primordial binaries with longer initial orbital periods undergo MT when the primary evolves into a giant star. This process results in an SS+MS system after envelope stripping. In the CE evolution channel, primordial binaries have the longest initial orbital periods. When the primary overfills its Roche lobe, MT becomes dynamically unstable due to the developed deep convective envelope, triggering a CE phase. If the binary survives, an SS+MS binary remains. All three channels ultimately lead to the formation of He+MS binaries. 
                \begin{figure*}[htp]
                    \centering
                    \includegraphics[width=0.8\textwidth]{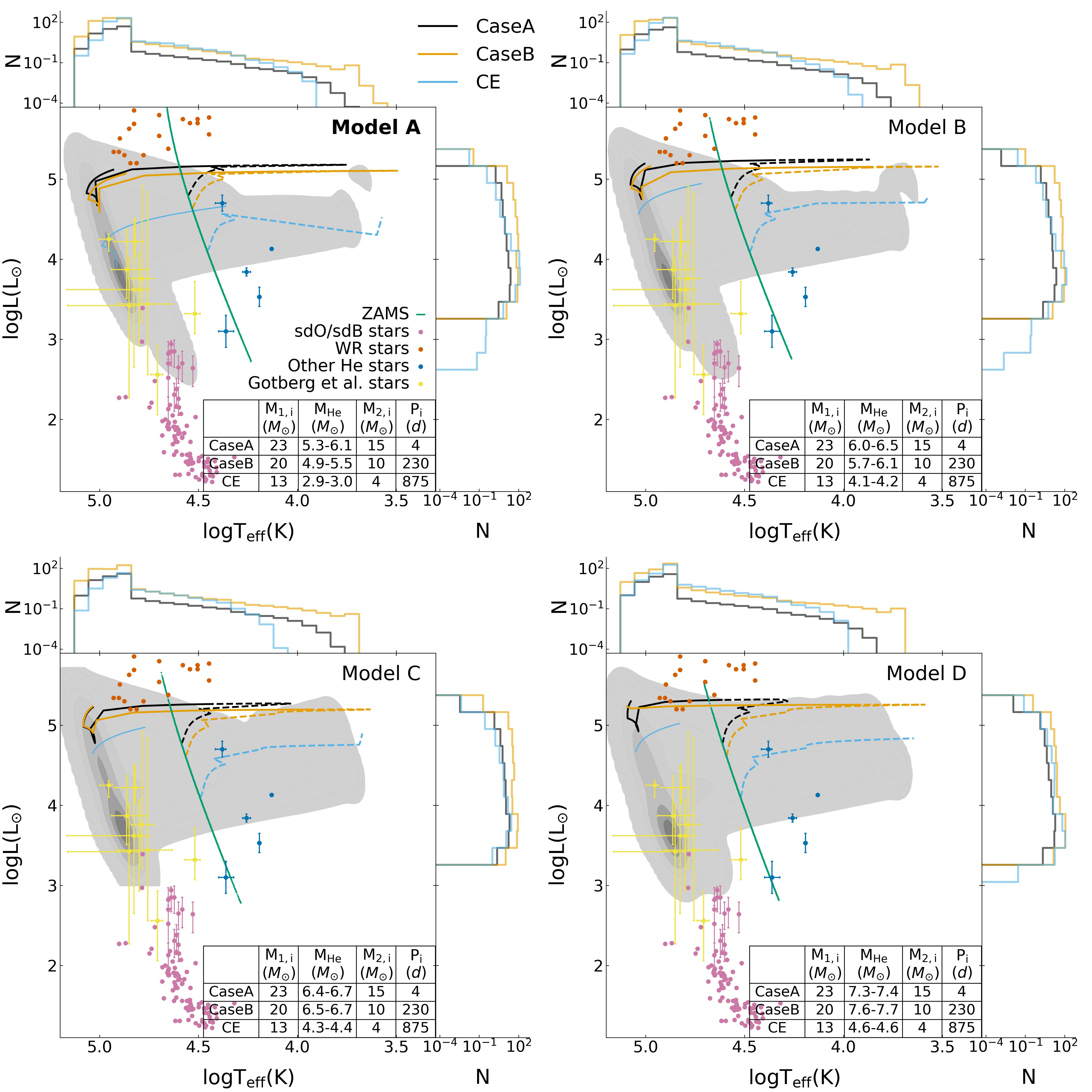}
                    \caption{Hertzsprung-Russell diagram showing the evolutionary tracks and number distributions (marginal panels) of IMHeS in He+MS binaries for four metallicity environments (Models A–D). Evolutionary tracks for individual binaries in the Case A MT (black), Case B MT (orange), and CE evolution (skyblue) channels are shown, with dashed and solid lines representing the progenitor phase before stripping and the subsequent IMHeS evolution (during which $M_{\mathrm{He}}$ decreases due to stellar wind mass loss), respectively. Observed populations are overlaid for comparison: WR stars \citep[vermillion circles,][]{Crowther2007,Hainich2014A&A...565A..27H}, hot subdwarfs \citep[reddish purple circles,][]{Lisker2005A&A...430..223L,Stroeer2007A&A...462..269S}, IMHeS \citep[yellow dots,][]{Gotberg2023ApJ...959..125G}, and additional candidate systems (blue dots). The bluish green band corresponds to the ZAMS.}
                        \label{fig: mergeHRDHe+MS_4}
                \end{figure*}

                \begin{figure}[htbp]
\centering
\includegraphics[width=0.4\textwidth]{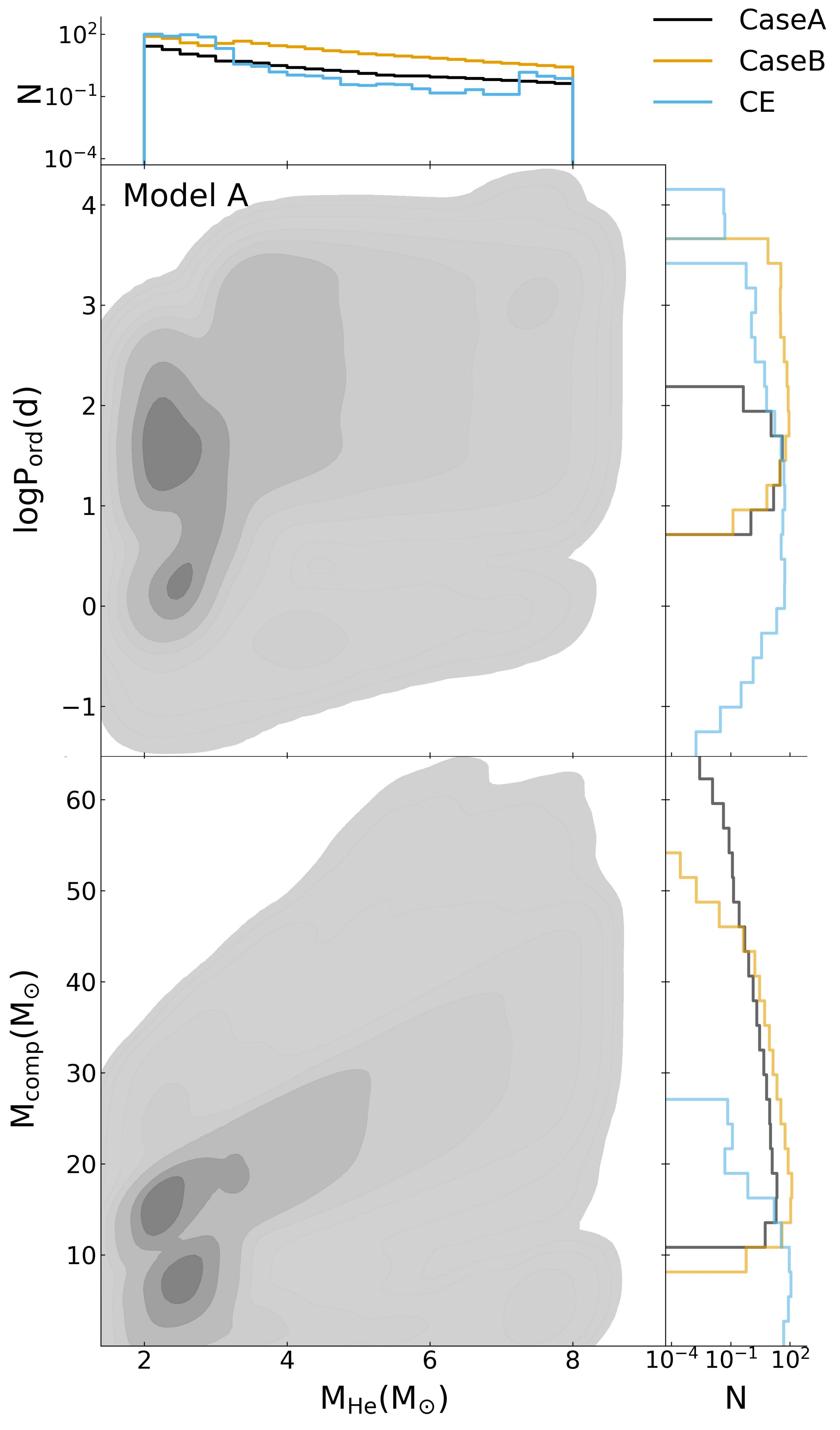}
                        \caption{Number distributions of He+MS binaries as functions of IMHeS mass and orbital period (top panel), binary masses (bottom panel) for the fiducial model (Model A). The marginal panels show the corresponding one-dimensional number distributions for systems evolving through the Case A MT (black), Case B MT (orange), and CE evolution (skyblue) channels. }
                        \label{fig: a1M1Z02w1MassPeriod_Mass2_He+MS}
            
                \end{figure}

Figure \ref{fig: mergeHRDHe+MS_4} presents the evolutionary tracks of IMHeS in He+MS binaries on the Hertzsprung-Russell diagram for Models A$-$D, with marginal panels showing number distributions. Black, orange, and skyblue lines denote the Case A MT, Case B MT, and CE evolution channels, respectively. Dashed lines represent the progenitor phase before stripping, while solid lines indicate the subsequent IMHeS evolution. Red dots with error bars mark observed IMHeS from \cite{Gotberg2023ApJ...959..125G}, and gray dots denote several candidate systems, including VFTS 291 \citep{McEvoy2015A&A...575A..70M}, LB-1 \citep{Shenar2020A&A...639L...6S}, $\gamma$ Columbae \citep{Irrgang2022}, SMCSGS-FS 69 \citep{Ramachandran2023A&A...674L..12R} and $\nu$ Sagittarii \citep{Gilkis2023MNRAS.518.3541G}. The gray shaded region encompasses all evolutionary tracks of IMHeS, with the highest density area corresponding to the HeMS band. The total numbers of He+MS systems in Models A$-$D range from $\sim 520$ to $\sim 950$ (see Table \ref{tab: Num_He}). A comparison of the results in Figure \ref{fig: mergeHRDHe+MS_4} indicates that IMHeS at different metallicities occupy similar regions in the Hertzsprung-Russell diagram (see Figure \ref{fig: mergeHRDHe+MS_6} for other models we simulate).

In the top panel of Figure \ref{fig: a1M1Z02w1MassPeriod_Mass2_He+MS}, we present the number distributions of He+MS binaries as a function of IMHeS mass ($M_{\mathrm{He}}$) and orbital period ($P_{\mathrm{orb}}$)\footnote{Throughout this work, the stellar and binary parameters presented in the figures (e.g., IMHeS mass $M_{\mathrm{He}}$, companion mass $M_{\mathrm{comp}}$, orbital period $P_{\mathrm{orb}}$) correspond to the current evolutionary state of each system during the IMHeS phase. We trace the full evolutionary history of each binary and record these parameters at each timestep. The population statistics are then constructed by weighting each timestep according to the system's lifetime and the assumed SFR.}, with marginal panels showing one-dimensional distributions. The orbital periods for He+MS binaries formed via Case A MT, Case B MT, and CE evolution span $\sim 3-100$ days, $\sim 5-3000$ days, and $\sim 0.1-2000$ days, respectively. 

The bottom panel of Figure \ref{fig: a1M1Z02w1MassPeriod_Mass2_He+MS} shows the number distributions as a function of the component masses of He+MS binaries. The companion mass ($M_{\mathrm{comp}}$) ranges are $\sim11-67\,M_{\odot}$, $\sim9-52 \,M_{\odot}$, and $\sim0.1-14\,M_{\odot}$ for systems evolved through the Case A MT, Case B MT, and CE evolution channels, respectively. Companion stars in binaries that previously experienced SMT can accrete more material, particularly in short-period systems evolving from a slow ($> 1 \mathrm{Myr}$) Case A MT phase \citep{Pols1994}. In contrast, systems surviving the CE phase do not allow accretion. Furthermore, progenitors with lower mass are more prone to CE evolution due to higher mass ratios, resulting in a substantial number of companions with masses below 8 $M_{\odot}$. 

                \begin{figure}[htbp]
\centering
\includegraphics[width=0.4\textwidth]{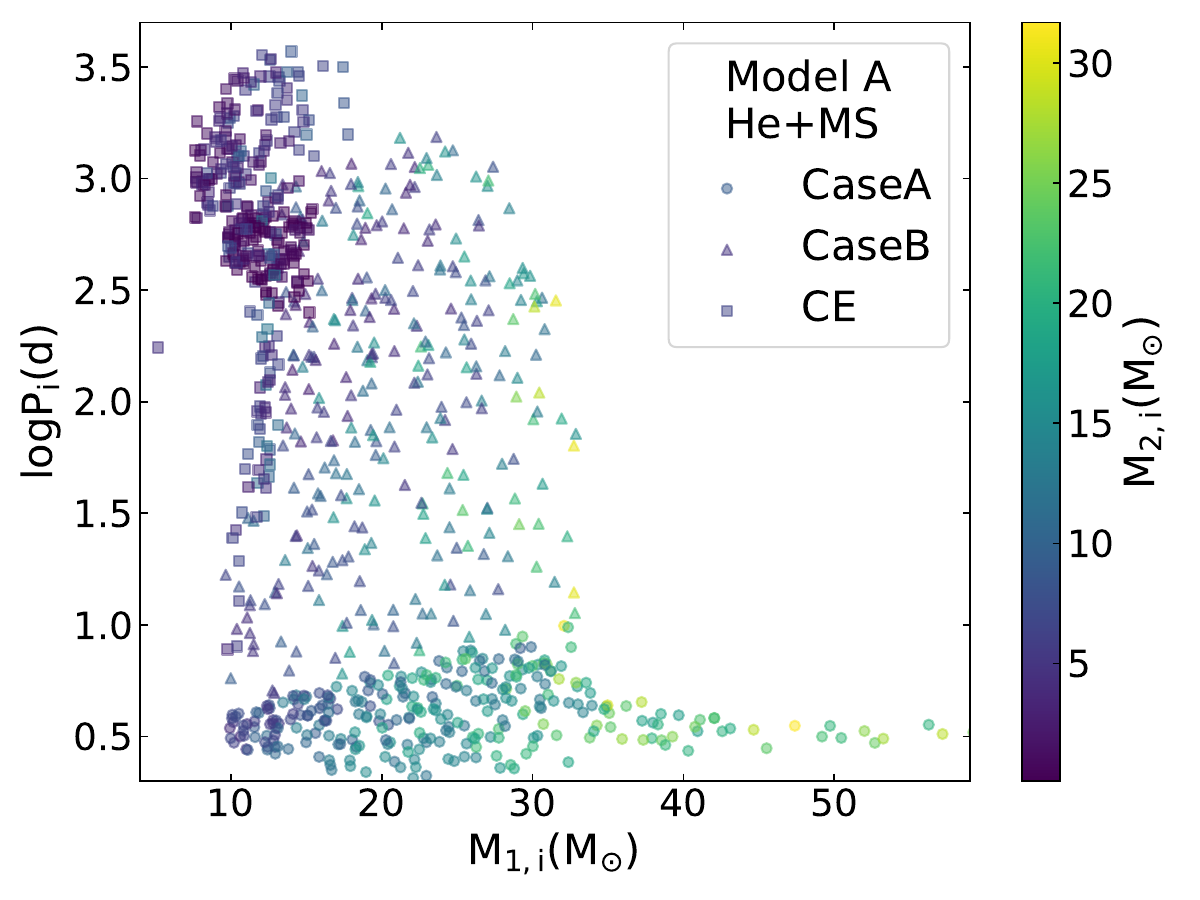}
                        \caption{Parameter distributions of primordial binaries that can evolve into He+MS binaries. Dots, triangles, and squares represent systems that will follow the Case A MT, Case B MT, and CE evolution channels, respectively. The color bar indicates the mass of the initial secondary star}.
                        \label{fig: a1M1Z02w1ZAMSmxmx2PHe+MS}

                \end{figure}

Figure \ref{fig: a1M1Z02w1ZAMSmxmx2PHe+MS} displays the parameter distribution of the primordial binaries that evolve into He+MS systems via the three channels. 
Primordial binaries that will undergo the Case A MT channel typically have initial orbital periods of $P_{\mathrm{orb}} \lesssim 10$\,days,  primary masses of $\sim 10-60\,M_{\odot}$ and secondary masses of $\sim 4-32\,M_{\odot}$. For primordial binaries that will undergo the Case B MT channel, their initial orbital periods range from $\sim 2$ days to $\sim 2000$ days, primary masses of $\sim 10$--$43\,M_{\odot}$, and secondary masses of $\sim 3$--$32\,M_{\odot}$. For primordial binaries that will undergo CE channel, initial primary masses are $\sim 5-27\,M_{\odot}$, initial secondary masses $\sim 0.1-14\,M_{\odot}$, and  initial orbital periods vary between $\sim 22$ days and $\sim 5000$\,days.

\subsubsection{He+G and He+He binaries}

\begin{figure}[htbp]
\centering
\includegraphics[width=0.4\textwidth]{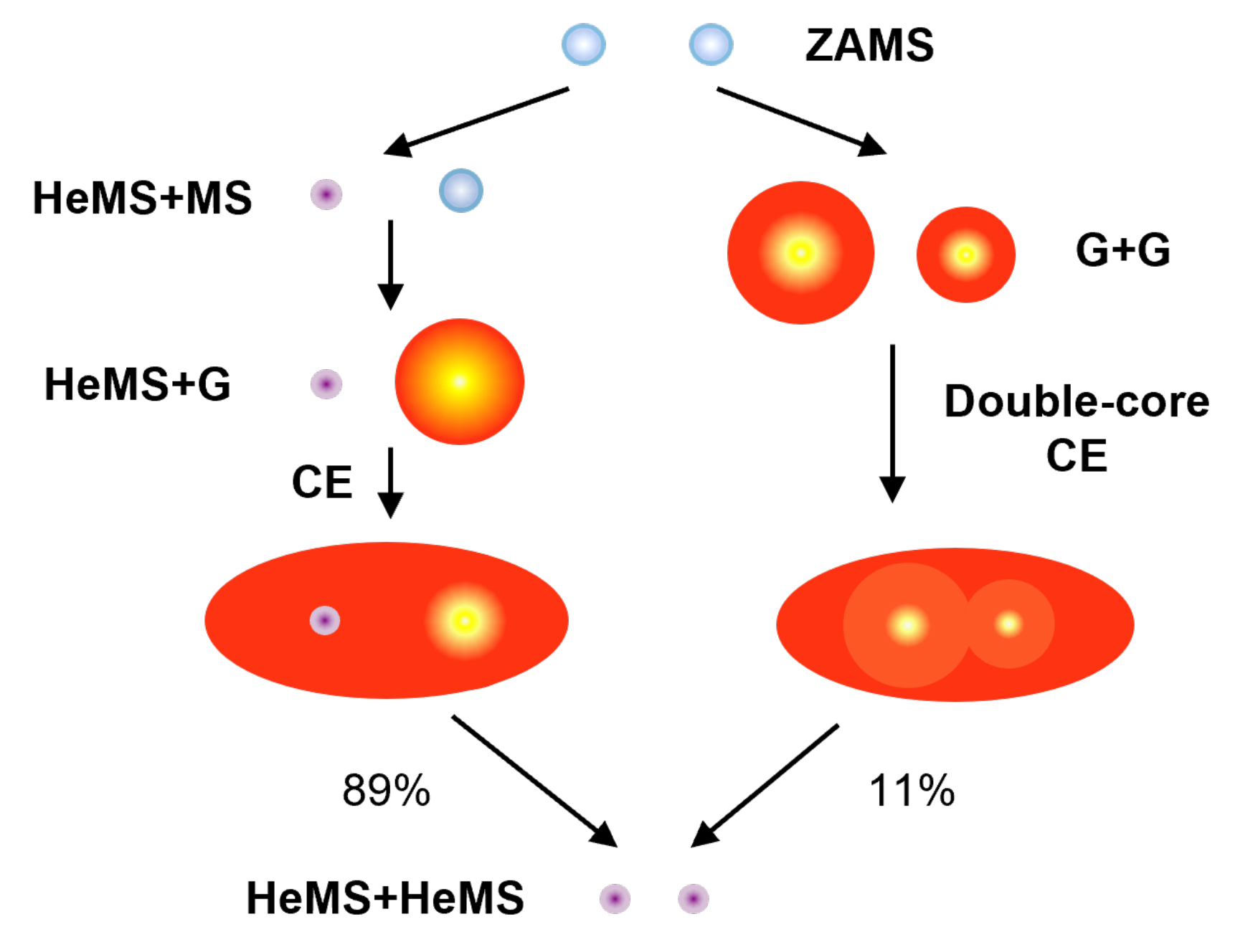}
                        \caption{Schematic illustration of the dominant formation channel for He+He binaries. These systems are primarily produced when a He+MS binary evolves, and the MS star expands into a giant, triggering a CE phase. A minority of systems may form via the double-core CE channel when both stars in a primordial binary evolve into giants simultaneously.}
                        \label{fig: He+He_evolutionaryChannel}
 
                \end{figure}

As He+MS binaries evolve, they give rise to two additional types of IMHeS systems: He+G and He+He binaries (Figure \ref{fig: He+He_evolutionaryChannel}). First, He+G systems form when the MS companion in an He+MS binary evolves into a giant. 
For all models, the number of He+G systems is below $\sim10$ (see Table \ref{tab: Num_He}) due to the short lifetime of giants.
The giant masses range from $2\,M_{\odot}$ to $30\,M_{\odot}$, peaking at $\sim 5\,M_{\odot}$. Given the extreme mass ratio, MT between the He star and the giant is almost always unstable ($> 96\%$), triggering a CE phase. This often leads to the formation of a second IMHeS and, consequently He+He binaries. The total number of He+He systems reaches $\sim 50$. These binaries can form via SMT+CE and CE+CE channels, depending on whether SMT or CE occurred during the evolution of primordial binaries.

An additional channel for forming He+He systems is the double-core CE channel \citep{Brown1995}. This occurs when both stars in a primordial binary evolve into giants, leading to CE evolution. If the CE is successfully ejected, a binary system consisting of two He cores remains. This channel requires similar component masses in the primordial binary and contributes only about $11\%$ of all He+He systems.

                \begin{figure}[htbp]
\centering
\includegraphics[width=0.4\textwidth]{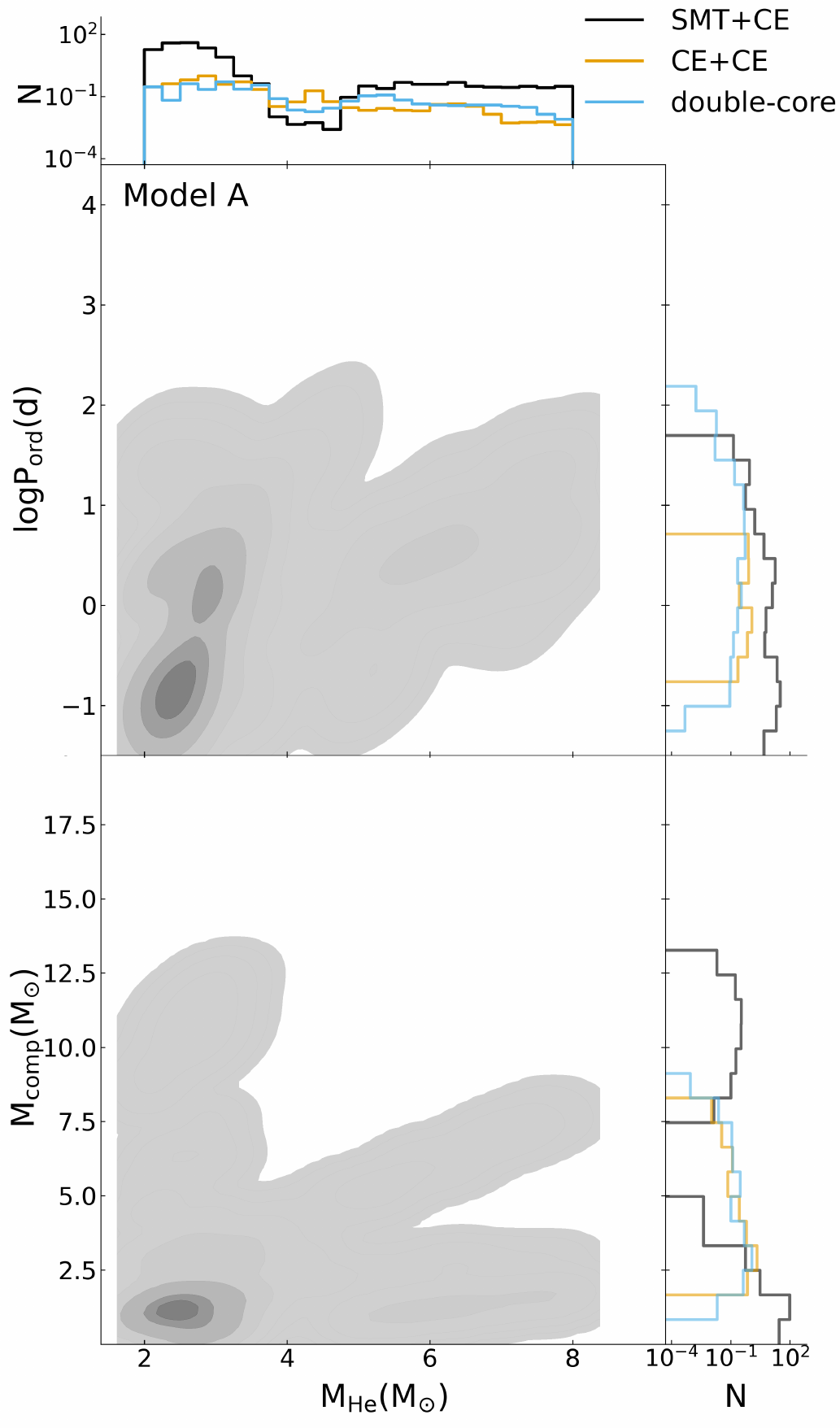}
                        \caption{Similar to Figure \ref{fig: a1M1Z02w1MassPeriod_Mass2_He+MS}, but for He+He binaries.}
                        \label{fig: a1M1Z02w1MassPeriod_Mass2_He+He}

                \end{figure}

                \begin{figure}[htbp]
\centering
\includegraphics[width=0.4\textwidth]{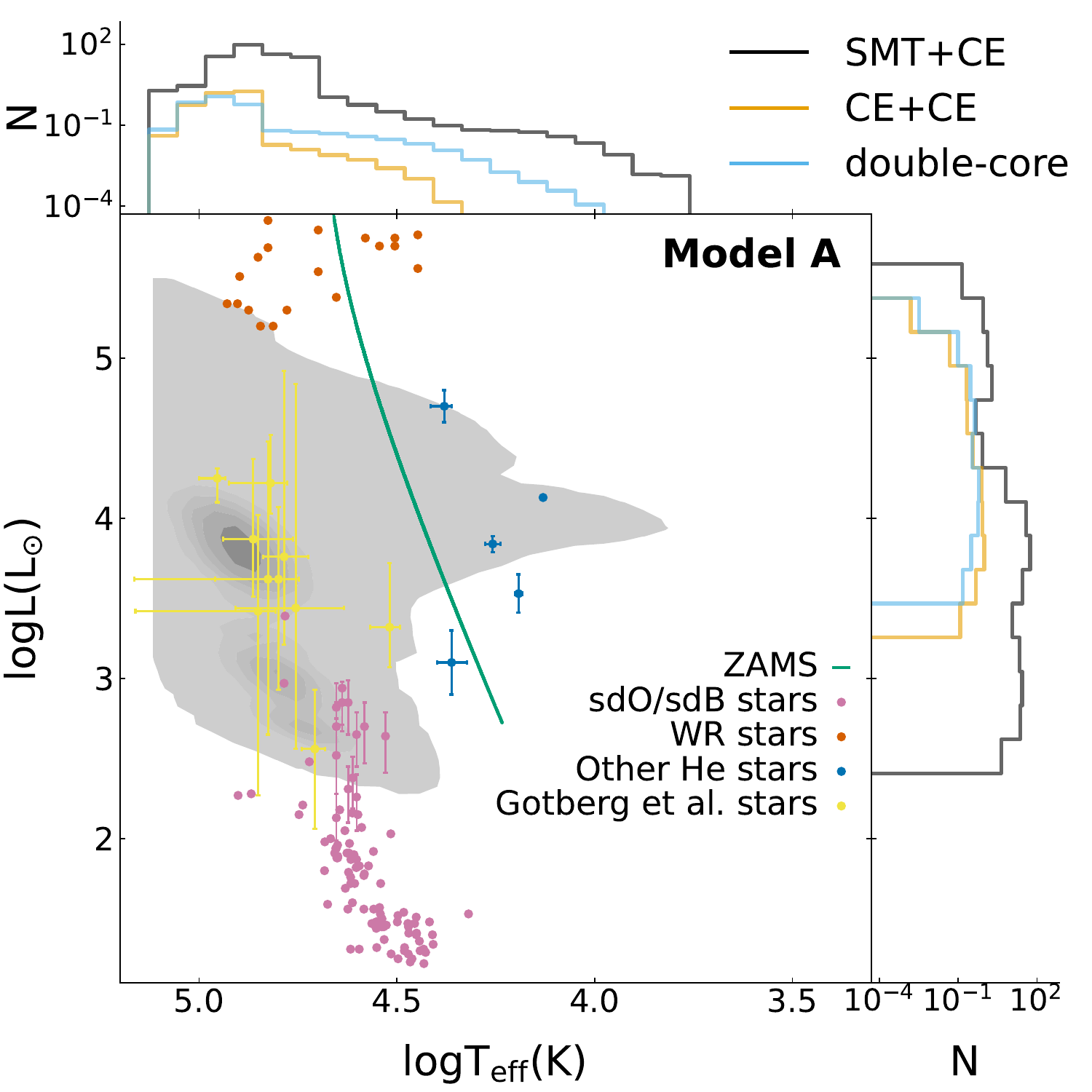}
                        \caption{Similar to Figure \ref{fig: mergeHRDHe+MS_4}, but for He+He binaries from our fiducial model.}
                        \label{fig: a1M1Z02w1HRDHe+He}

                \end{figure}

The top panel of Figure \ref{fig: a1M1Z02w1MassPeriod_Mass2_He+He} shows the number distributions of He+He binaries as a function of IMHeS mass ($M_{\mathrm{He}}$) and orbital period ($P_{\mathrm{orb}}$). The orbital periods range from $\sim0.05$ days to $\sim100$ days, significantly shorter than those of He+MS systems, as CE evolution efficiently shrinks the orbital separation. Notably, He+He binaries experiencing two CE phases are rare, contributing $\sim 10\%$ of the total population, because post-CE systems are prone to merging during the second CE phase.

The bottom panel of Figure \ref{fig: a1M1Z02w1MassPeriod_Mass2_He+He} shows the number distributions as a function of the component masses of He+He binaries. The companion mass ($M_{\mathrm{comp}}$) spans $\sim 0.2-14\, M_{\odot}$ and $\sim 2-7\,M_{\odot}$ for systems formed through the SMT+CE and CE+CE channels, respectively. Systems from the SMT+CE channel exhibit a broader mass range. This can be attributed to two main factors.
On one hand, companion stars undergoing mass loss during CE evolve into He stars with masses below $2\, M_{\odot}$, giving rise to IMHeS+sdOB binaries. On the other hand, CE evolution substantially reduces both the orbital periods and total masses of binary systems, thereby inhibiting the formation of IMHeS+WR systems. As a result, IMHeS+WR systems are exclusively formed through the SMT+CE channel.

The Hertzsprung-Russell diagram in Figure \ref{fig: a1M1Z02w1HRDHe+He} shows that the IMHeS in He+He systems are generally located near the HeMS band. 
This is because He+He binaries, after undergoing a CE phase, possess relatively short orbital periods, and their subsequent evolution during the He giant phase is restricted by the Roche lobe. 
A notable difference for the He+He population, compared to other IMHeS populations, is the scarcity of HeG stars with large radii. 

\subsubsection{He+CO binaries}

                \begin{figure}[htbp]
\centering
\includegraphics[width=0.4\textwidth]{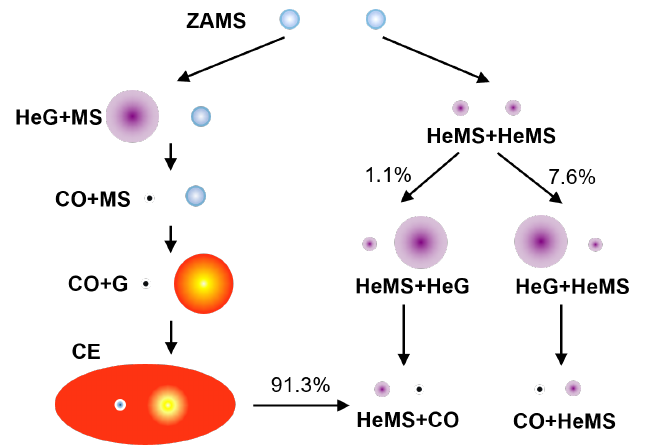}
\caption{Formation channels for He+CO binaries. Most systems originate from a CO+MS binary that undergoes a CE phase, while a smaller fraction descend from a He+He binary after one component explodes in an SN.}
\label{fig: He+CO_evolutionaryChannel}

\end{figure}

As the descendants of He+MS binaries, systems consisting of an IMHeS and a CO (i.e., He+CO binaries) constitute another important category of the IMHeS population. 
Figure \ref{fig: He+CO_evolutionaryChannel} illustrates the evolutionary pathways for He+CO binaries. Firstly, the IMHeS in a He+MS binary evolves into a CO, resulting in a CO+MS binary. When the MS star ascends the giant branch, the system enters a CE phase due to the extreme mass ratio, ultimately forming an He+CO binary. Alternatively, the more massive IMHeS in an He+He binary can evolve into the He giant phase and undergo an SN, leaving a CO companion. 

                \begin{figure}[htbp]
\centering
\includegraphics[width=0.4\textwidth]{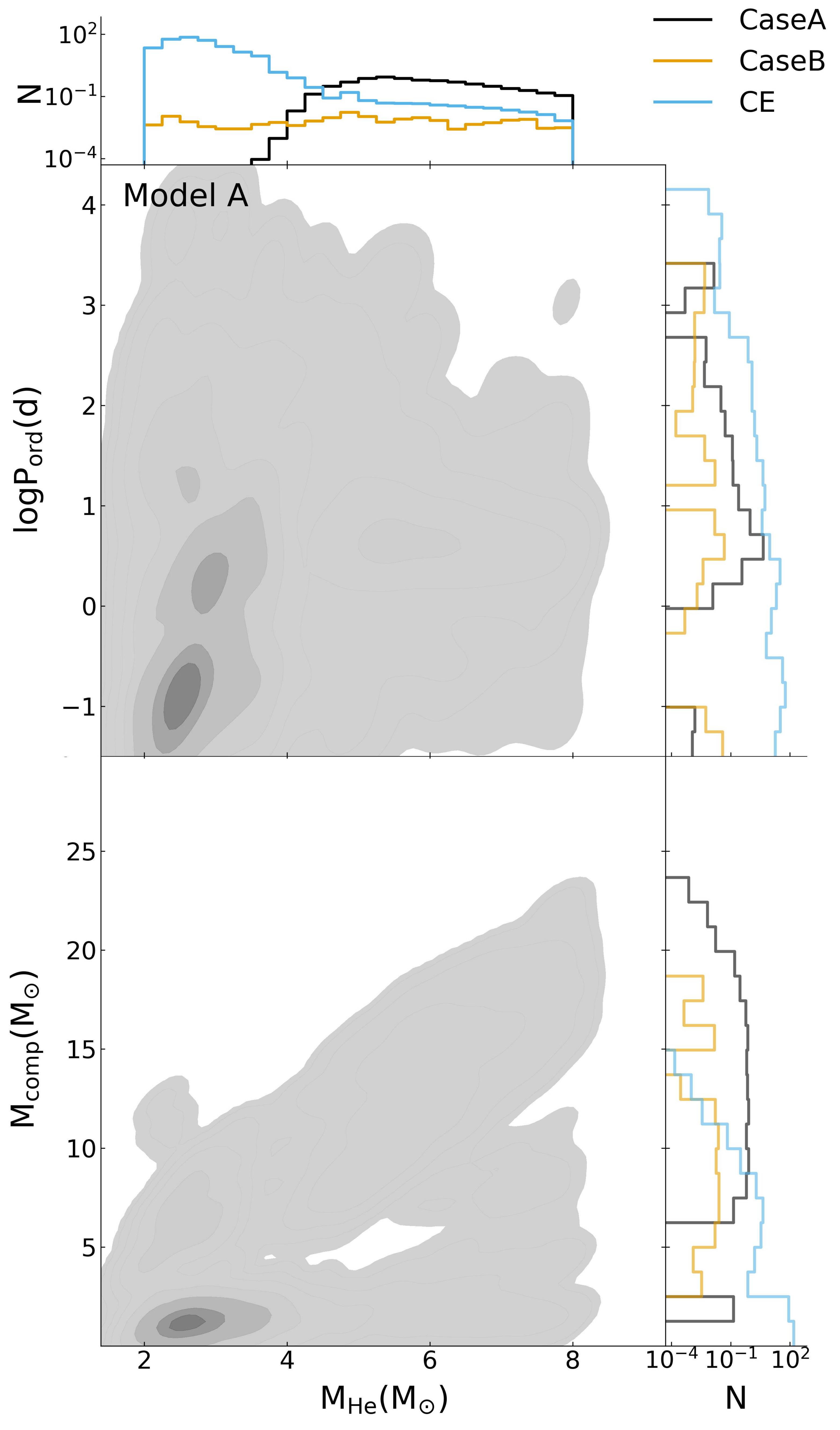}
                        \caption{Similar to Figure \ref{fig: a1M1Z02w1MassPeriod_Mass2_He+MS}, but for He+CO binaries.}
                        \label{fig: a1M1Z02w1MassPeriod_Mass2_He+CO}

                \end{figure}

In the top panel of Figure \ref{fig: a1M1Z02w1MassPeriod_Mass2_He+CO}, we present the number distributions of He+CO binaries as a function of IMHeS mass ($M_{\mathrm{He}}$) and orbital period ($P_{\mathrm{orb}}$). The orbital periods cover a wide range of $\sim 0.05-10000$ days, primarily due to changes in the orbits caused by SN kicks. 

                \begin{figure}[htbp]
            \centering
            \includegraphics[width=0.4\textwidth]{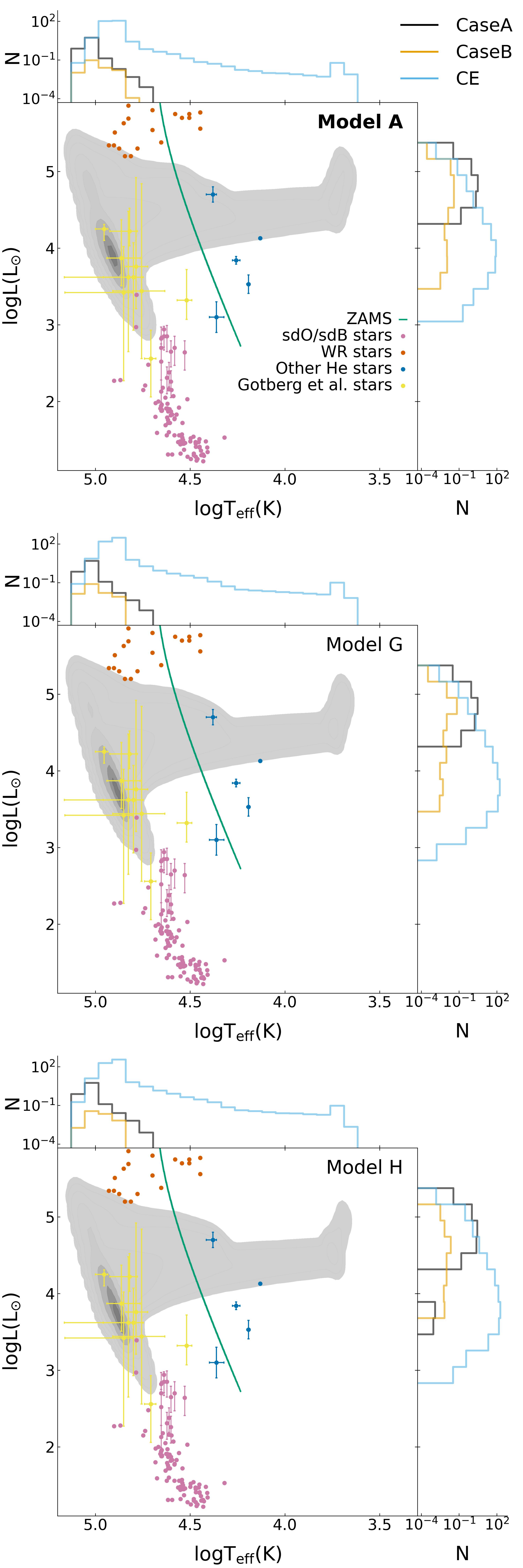}
                        \caption{Hertzsprung-Russell diagram of the He+CO binary population for different CE ejection efficiencies $\alpha_{\mathrm{CE}}$. Models A, G, and H correspond to $\alpha_{\mathrm{CE}}=1$, 3 and 5, respectively.}
                        \label{fig: mergeHRDHe+CO_3}
                \end{figure}

The bottom panel of Figure \ref{fig: a1M1Z02w1MassPeriod_Mass2_He+CO} shows the number distribution as a function of the component masses of He+CO binaries. The companion mass ($M_{\mathrm{comp}}$) ranges are $\sim 0.2-22\, M_{\odot}$, $\sim3-15\,M_{\odot}$ and $\sim 0.2-17 \,M_{\odot}$ for systems evolved through Case A MT, Case B MT and CE evolution channels, respectively. 
Only the Case A MT channel can produce binaries with $M_{\mathrm{CO}} > 17\,M_{\odot}$. 
Based on final masses, CO companions are classified as WDs ($\sim 0.6-1.4\,M_{\odot}$), NSs ($\sim1.1-2.5\,M_{\odot}$), and BHs ($\sim2.5-23\,M_{\odot}$). He+WD and He+NS binaries form predominantly via the CE evolution channel, while He+BH binaries are formed through both the SMT and CE evolution channels, with each channel contributing significantly.

The Hertzsprung-Russell diagram in Figure \ref{fig: mergeHRDHe+CO_3} shows the populations for our fiducial model and for Models G and H (with $\alpha_{\mathrm{CE}} = 3$ and 5, respectively). 
We can see that different CE ejection efficiencies have a relatively minor impact on the properties of He+CO binaries but significantly affect their population numbers.
From Table \ref{tab: Num_He}, 
the number of He+CO binaries increases from $\sim 120$ to $\sim 650$ in Model G and $\sim 750$ in Model F. This enhancement is primarily caused by He+WD systems, which increase by more than an order of magnitude (from $\sim 30$ in the fiducial model to $\sim 550$ in Model G and $\sim 620$ in Model H) and dominate the overall growth of the He+CO population.

\subsubsection{Single IMHeS} \label{sec: Single IMHeS}

                \begin{figure}[htbp]
\centering
\includegraphics[width=0.45\textwidth]{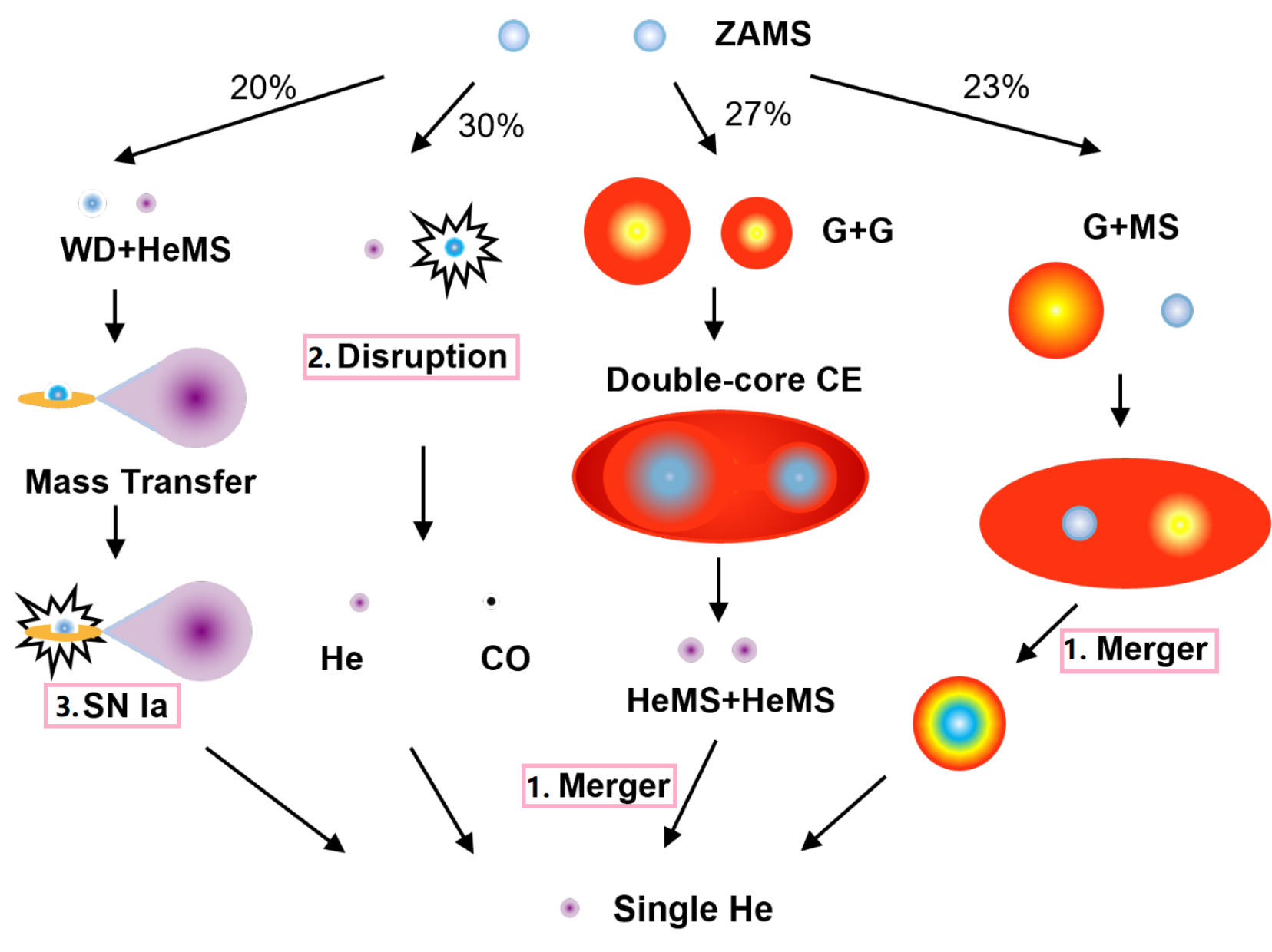}
                        \caption{Formation channels for single IMHeS. These include: (1) the merger channel, where a binary merges during a failed CE ejection or through the merger of two He stars; (2) the Type Ia SN channel, where a thermonuclear explosion of a WD disrupts the binary; and (3) the disruption channel, where an asymmetric core-collapse SN imparts a natal kick that unbinds the system.} 
                        \label{fig: SingleHe_evol}

                \end{figure}
                \begin{figure}[htbp]
\centering
\includegraphics[width=0.4\textwidth]{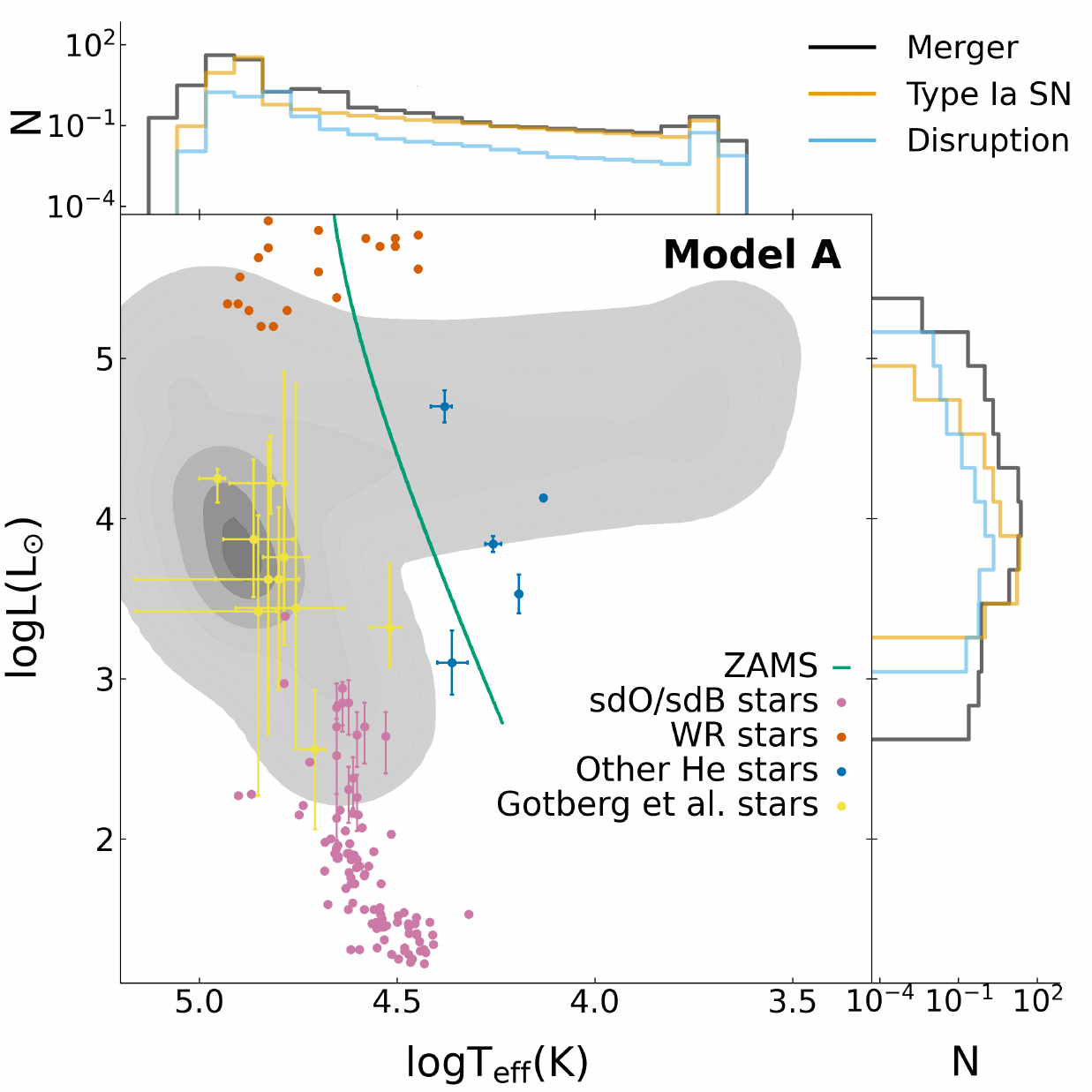}
                        \caption{Similar to model A in Figure \ref{fig: mergeHRDHe+MS_4}, but for single IMHeS originating from three distinct formation channels.}
                        \label{fig: a1M1Z02w1HRDHe}
   
                \end{figure}

                \begin{figure}[htbp]
\centering
\includegraphics[width=0.4\textwidth]{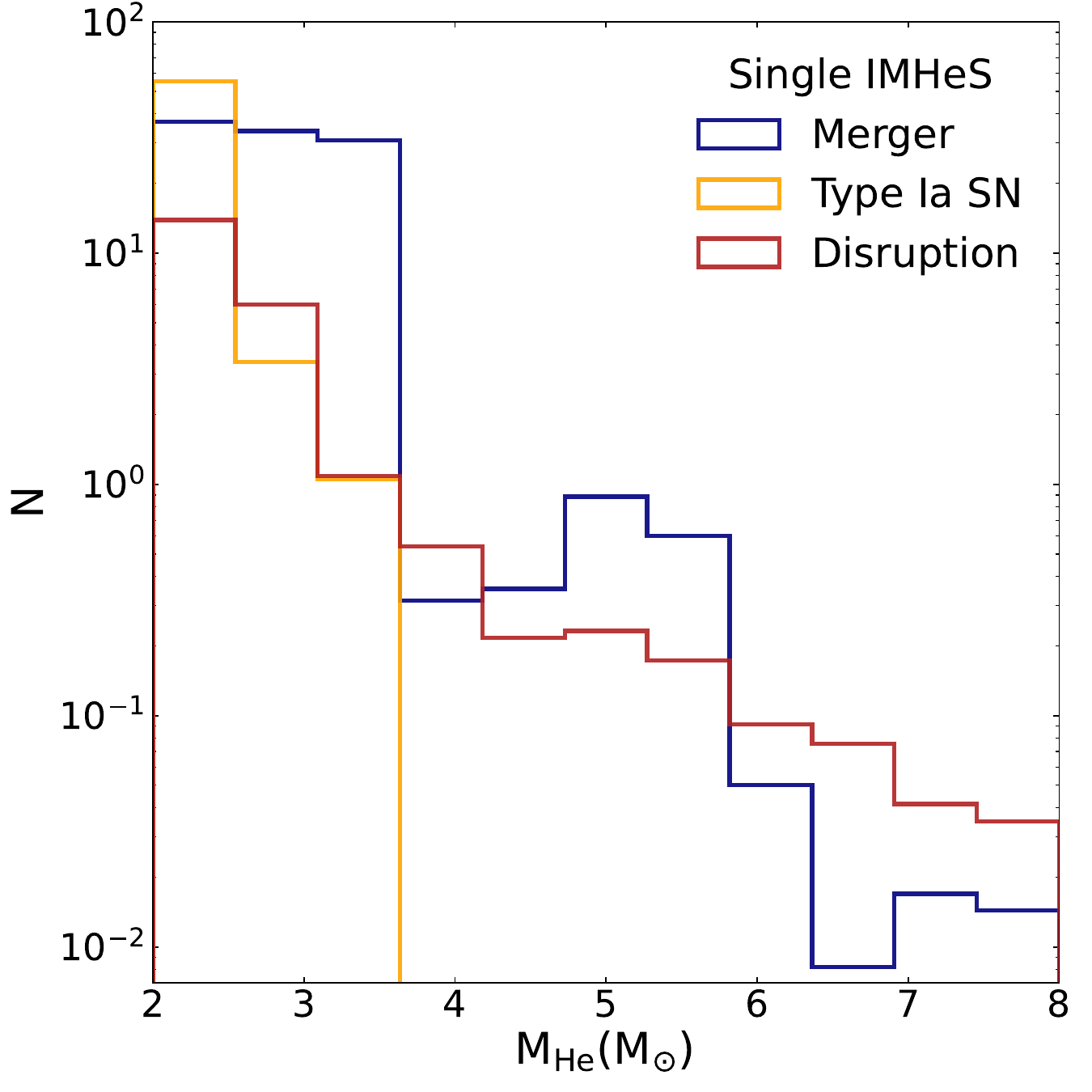}
                        \caption{Mass distribution of single IMHeS from our fiducial model.}
                        \label{fig: a1M1Z02w1SHeSMdistribution}
         
                \end{figure}
The Single IMHeS discussed here refers to isolated IMHeS that have undergone stripping of their outer hydrogen envelope through binary interactions. 
Figure \ref{fig: SingleHe_evol} presents three main formation channels: 

(1) Merger channel: A G+MS binary may initiate a CE phase. If the envelope ejection fails, the binary merges within the CE, potentially forming a single IMHeS. Alternatively, the merger of two He stars can also lead to a single IMHeS.

(2) Type Ia SN channel: In close WD+He systems, MT from the IMHeS to a carbon-oxygen WD companion can induce a type Ia SN, disrupting the system and leaving an isolated IMHeS. 

(3) Disruption channel: An asymmetric core-collapse SN imparts a natal kick to the newly formed CO. If the kick velocity is sufficiently large, the binary can be disrupted, ejecting the companion and leaving a single IMHeS.

These three channels produce single IMHeS with slightly different properties on the Hertzsprung-Russell diagram (see Figure \ref{fig: a1M1Z02w1HRDHe}), with each channel contributing significantly to the overall population. The merger and disruption channels can produce single IMHeS with masses  up to $8\, M_{\odot}$, whereas the Type Ia SN channel is truncated at $\sim 3-4\, M_{\odot}$. This truncation arises because successful Type Ia events require SMT from the IMHeS to the WD, which is only possible for low-mass ($\lesssim 4\,M_\odot$) donors; higher-mass donors would lead to dynamically unstable MT.
As shown in Figure \ref{fig: a1M1Z02w1SHeSMdistribution}, the mass distribution of single IMHeS is predominantly at the low-mass end, with only a few IMHeS having masses $\gtrsim 4\,M_{\odot}$.

\section{Discussion and Conclusion} \label{sec: conclusion}

The search for IMHeS has gained increasing attention, yet their theoretical formation pathways remain inadequately understood. While recent observational studies provide compelling evidence for their existence \citep{Drout2023Sci...382.1287D}, their detection is often hindered by the presence of luminous companions.

To investigate the formation and properties of IMHeS, we employed a binary population synthesis approach under a range of physical assumptions. Our statistical analysis focused on five main categories of IMHeS populations: He+MS, He+G, He+He, He+CO, and single IMHeS. 
Among these, He+MS systems constitute the dominant population, accounting for $\gtrsim 50\%$ of all IMHeS (see Table \ref{tab: Num_He}). These systems form through three channels: Case A MT, Case B MT and CE evolution channels. In contrast, He+G and He+He systems have relatively short evolutionary timescales, resulting in comparatively small predicted populations ($\sim 10$ and $\sim 50$ systems, respectively, in the fiducial model).
The evolutionary pathways of the He+CO population are complex. Their predicted numbers are most sensitive to the options of CE ejection efficiency. Furthermore, He+CO binaries exhibit an extremely broad range of orbital periods ($\sim 0.05 - 10000 $ days).
Single IMHeS are relatively rare, forming mainly through binary mergers or SN-induced disruptions channels that can explain systems such as $\gamma$ Columbae.

Despite the predicted abundance of He+MS systems, a key question is why so many He+MS systems remain observationally unidentified. We attribute this to two main factors. First, residual hydrogen-rich envelopes may persist following either Case B MT or CE evolution \citep{Nie2025ApJ}, complicating the spectral identification of IMHeS. Second, the majority of He+MS systems contain massive O- or B-type companions whose luminosities significantly outshine those of IMHeS. In such cases, the spectral features of the He star become too weak to discern, rendering conventional spectral analysis ineffective for confirming the presence of an IMHeS \citep{Drout2023Sci...382.1287D,Ludwig2026}.

We further modeled the populations of IMHeS in the Milky Way and nearby low-metallicity galaxies such as the LMC and SMC. Adopting SFRs of 2 $M_{\odot} \,\mathrm{yr^{-1}}$ for the Milky Way \citep{Licquia2015ApJ...806...96L}, 0.2 $M_{\odot} \,\mathrm{yr^{-1}}$ for the LMC \citep{Harris2009AJ....138.1243H} and 0.05 $M_{\odot}\, \mathrm{yr^{-1}}$ for the SMC \citep{Rubele2015}, we estimate  $\sim 2360$ IMHeS in the Milky Way using our solar-metallicity model, $\sim 230$ in the LMC using our LMC-like metallicity model, and $\sim 30$ in the SMC using our SMC-like metallicity model.

Our predicted IMHeS numbers are lower than recent estimates from \citet{Yungelson2024A&A...683A..37Y}, \citet{Hovis-Afflerbach2025A&A...697A.239H}, and \citet{Blomberg2026}. This discrepancy arises from a combination of differences in modeling approaches, input physics, and population synthesis methodologies. A fundamental difference lies in the stellar evolution codes employed. We use the rapid binary evolution code BSE \citep{Hurley2002}, whereas the aforementioned studies utilize the detailed stellar evolution code MESA \citep{Paxton2011}. Direct comparisons are complicated by inherent differences in the treatment of stellar interior physics, including convective and semiconvective mixing efficiencies, as well as the response of stellar structure to mass exchange.

\citet{Yungelson2024A&A...683A..37Y} focused exclusively on IMHeS formed through SMT and predicted $\sim3000$ He+MS systems in the Milky Way, assuming an SFR of 2 $M_{\odot} \,\mathrm{yr^{-1}}$ and a binary fraction of 50\%. In contrast, our fiducial model yields $\sim 1200$ such systems under the same SFR but assuming a binary fraction of 100\%. We find that the choice of the initial mass function (IMF, $dN/dM \propto M^{-\alpha}$) is a primary driver of this discrepancy. \citet{Yungelson2024A&A...683A..37Y} adopted the Salpeter IMF with a slope of $\alpha = 2.3$ \citep{Salpeter1955}, whereas we employ the \citet{Kroupa1993} IMF with a steeper slope of $\alpha = 2.7$ \citep[see the Method section of][]{Shao2021b}. When we renormalize our models to adopt the Salpeter IMF, the predicted number of IMHeS increases by a factor of $\sim 3$, bringing it into closer agreement with the estimates of \citet{Yungelson2024A&A...683A..37Y}.

\citet{Hovis-Afflerbach2025A&A...697A.239H} and \citet{Blomberg2026} considered both SMT and CE formation channels. Although they do not explicitly quote the number of IMHeS binaries with $M_{\mathrm{He}}=2-8\,M_\odot$, their results can be estimated from the published figures. For an SFR of 2 $M_{\odot} \,\mathrm{yr^{-1}}$, these estimates imply a population of $\sim 6000$ such binaries, which is notably larger than our fiducial predictions. A key contributing factor is again the IMF; both studies adopted the \citet{Kroupa2001} IMF with a slope of $\alpha = 2.35$. Beyond the IMF, differences in physical assumptions further complicate direct comparisons. Our models incorporate distinct prescriptions for MT efficiency, MT stability, and CE evolution. These choices can significantly influence the fraction of binary systems that successfully produce IMHeS. Given the complexity of these processes, the individual and combined effects of such modeling differences are not straightforward to disentangle.

Our study also evaluated the influence of various physical models (including metallicity, MT schemes, CE evolution, and He star wind prescriptions) on the IMHeS population (Figure \ref{fig: merge_massdistribution_N_rate}). We find that metallicity and CE ejection efficiency are the dominant factors, both of which can affect the numbers of IMHeS by a factor of $\sim 2$. In contrast, other parameters such as MT efficiency and He star wind scaling factor have comparatively minor effects on the overall properties of IMHeS. 

Our predictions for the IMHeS population are subject to several sources of uncertainty.
The physics of SN explosions still remains unclear, the magnitude and direction of SN kicks \citep{Valli2025} imparted to NSs and BHs  can determine whether a binary survives or is disrupted, thereby modulating the corresponding IMHeS numbers. The sample size in Monte Carlo simulations introduces statistical fluctuations, particularly for rare evolutionary subchannels.
In our simulations, we evolve $10^6$ primordial binaries for each model, recording detailed evolutionary information. This level of detail results in a computational cost of approximately 1$–$2 days per model. While evolving $10^7 $ binaries per model would be computationally prohibitive and generate unmanageable data volumes, we have performed convergence tests by running models with $10^5$ binaries. The total number of IMHeS varies by less than $5\%$ between the $10^5$ to $10^6$ realizations, confirming that our simulations are statistically robust.

In summary, this work provides a comprehensive population synthesis analysis of IMHeS, elucidating their dominant formation channels, predicted abundances, and the physical parameters that govern their populations. Our results highlight the critical role of binary evolution in producing this elusive class of objects and offer testable predictions for future observational campaigns aimed at uncovering these hidden stellar populations.

\vspace{20pt}
We thank the anonymous referee for constructive suggestions that helped improve this paper. This work was supported by the National Key Research and Development Program of China (Grant Nos 2023YFA1607902 and 2021YFA0718500), and the Natural Science Foundation of China (Nos 12041301, 12121003, and 12373034). The following publicly available software packages were utilized in this work: \texttt{MATPLOTLIB} \citep{Matplotlib2007CSE.....9...90H}, \texttt{VAEX} \citep{Vaex2018A&A...618A..13B}, \texttt{NUMPY} \citep{numpy2020Natur.585..357H}, \texttt{SEABORN} \citep{Waskom2021zndo....592845W}, \texttt{PANDAS} \citep{pandas2022zndo...3509134T}. All input files to reproduce our results are available for download from Zenodo at \dataset[doi:10.5281/zenodo.19014072]{https://doi.org/10.5281/zenodo.19014072}.

\bibliography{article}{}
\bibliographystyle{aasjournal}

{\let\clearpage\relax
\appendix
}

        \renewcommand{\thefigure}{A.\arabic{figure}}
        \renewcommand{\thetable}{A.\arabic{table}}
        \setcounter{figure}{0}
        \setcounter{table}{0}

\begin{figure*}[htp]
            \centering
            \includegraphics[ height=0.8\textheight, keepaspectratio]{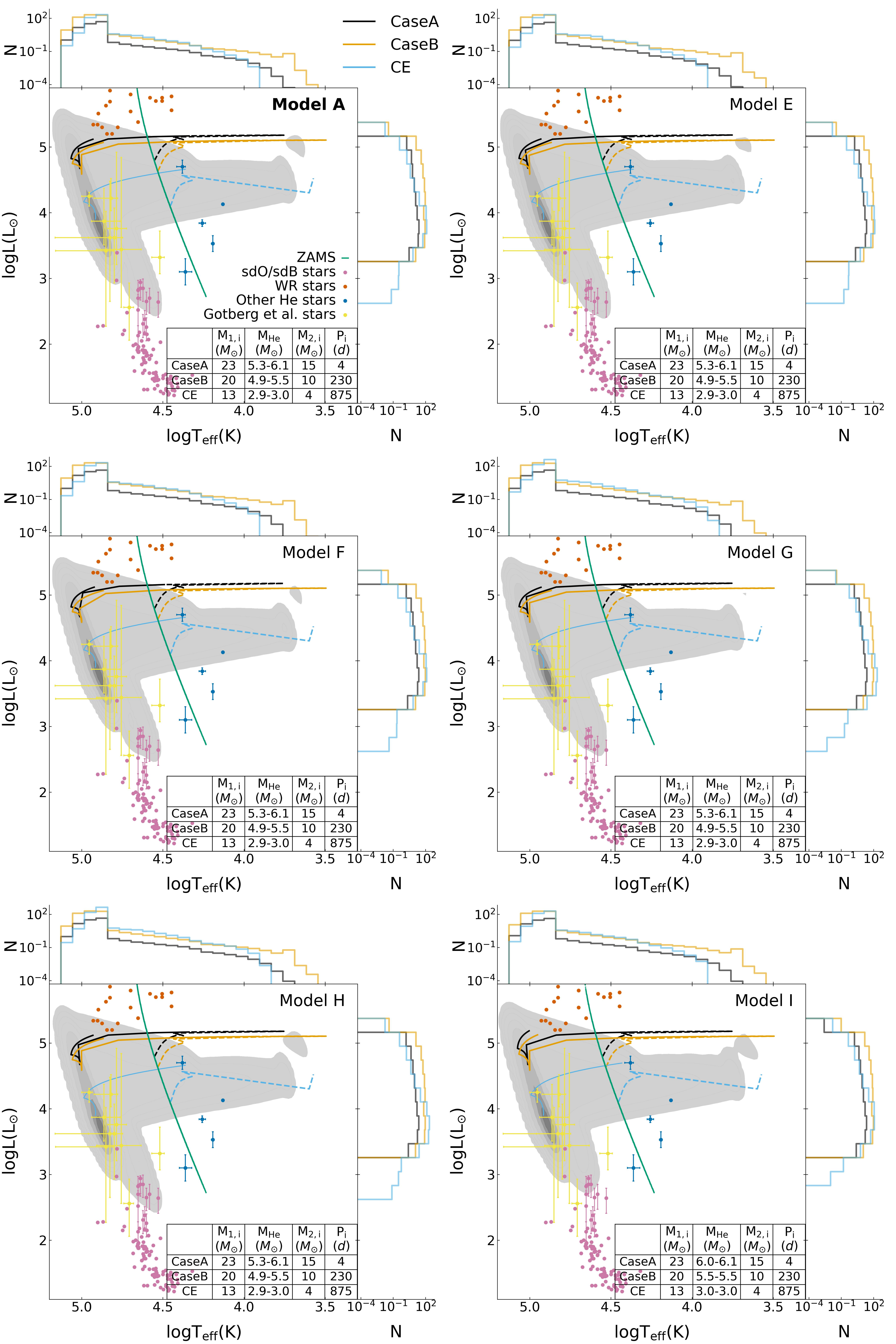}
            \caption{Similar to Hertzsprung-Russell diagram in Figure \ref{fig: mergeHRDHe+MS_4}, but for the He+MS populations under different physical assumptions. Models A, E, and F represent different MT schemes; Models A, G, and H correspond to CE ejection efficiencies of $\alpha_{\mathrm{CE}}=1$, 3, and 5; and Models A and I compare standard versus reduced He star wind prescriptions.}
            \label{fig: mergeHRDHe+MS_6}
        \end{figure*}

In this appendix, Figure \ref{fig: mergeHRDHe+MS_6} presents the Hertzsprung-Russell diagram distribution of the He+MS population under different model assumptions. Models A, E, and F represent different MT schemes; Models A, G, and H correspond to $\alpha_{CE}$ values of 1, 3, and 5, respectively; and Models A and I compare the standard and reduced wind scaling factors for helium stars. The formation of the He+MS population occurs early in the evolutionary sequence, rendering it largely insensitive to subsequent physical processes such as MT efficiencies, CE evolution, or wind prescriptions. This explains the relatively minimal variation observed across these model configurations.

\end{document}